\documentclass[floatfix,aps,twocolumn,prl,superscriptaddress, amsfonts]{revtex4}

\usepackage{graphicx}
\usepackage{dcolumn}
\usepackage{natbib}
\usepackage{bm}

\begin{document}
\title{Optical study of interactions in a d-electron Kondo lattice with ferromagnetism}
\author{K.S.~Burch}
\email{burch@physics.ucsd.edu}
\affiliation{Department of Physics, University of California, San Diego, CA 92093-0319}
\author{A.~Schafgans}
\affiliation{Department of Physics, University of California, San Diego, CA 92093-0319}
\author{N.P.~Butch}
\affiliation{Department of Physics and Institute for Pure and Applied Physical Sciences, University of California, San Diego, La Jolla, California 92093}
\author{T.A.~Sayles}
\affiliation{Department of Physics and Institute for Pure and Applied Physical Sciences, University of California, San Diego, La Jolla, California 92093}
\author{M.B.~Maple}
\affiliation{Department of Physics and Institute for Pure and Applied Physical Sciences, University of California, San Diego,
La Jolla, California 92093}
\author{B.C.~Sales}
\affiliation{Condensed Matter Sciences Division, Oak Ridge National Laboratory, Oak Ridge, Tennessee 37831}
\author{D.~Mandrus}
\affiliation{Condensed Matter Sciences Division, Oak Ridge National Laboratory, Oak Ridge, Tennessee 37831}
\author{D.N.~Basov}
\affiliation{Department of Physics, University of California, San Diego, CA 92093-0319}

\begin{abstract}
	We report on a comprehensive optical, transport and thermodynamic study of the Zintl compound Yb$_{14}$MnSb$_{11}$, demonstrating that it is the first ferromagnetic Kondo lattice compound in the underscreened limit. We propose a scenerio whereby the combination of Kondo and Jahn-Teller effects provides a consistent explanation of both transport and optical data.
\end{abstract}

\maketitle
	The past two decades have seen an enormous effort exerted to understand systems with strongly correlated electrons. In these materials, multiple interactions result in a litany of new behavior. Interestingly, some of these interactions compete such that they are rarely found together, but when they do coexist, the competition often produces a rich phase diagram.  For example in the manganites competition between Jahn-Teller and double exchange interactions leads to the phenomenon of colossal magneto-resistance.\cite{manginites} Kondo lattices are another example of materials where a variety of ground states  are exhibited, including Fermi liquids with heavy quasi-particles  and Kondo-insulators.\cite{sasa,hfoptics,coleman} Yet there are few materials where ferromagnetism and a Kondo-resonance can be found simultaneously,\cite{fmkondo} and none have been shown to exist in the underscreened limit (namely the local moment is larger than the number of screening channels) or to exhibit Jahn-Teller and Kondo effects. Furthermore, despite the large body of work on Kondo lattices, very few d-electron systems have been discovered that manifest heavy-fermion behavior.\cite{LiVO} 
	
	In this letter, we demonstrate that Yb$_{14}$MnSb$_{11}$ is the first underscreened, ferromagnetic, d-electron heavy fermion compound as well as the first to display Kondo and Jahn-Teller effects. The compound Yb$_{14}$MnSb$_{11}$ belongs to the "14-1-11" class of materials, which possess the chemical formula A$_{14}$MnPn$_{11}$, where A is an alkaline or rare earth atom and Pn is a Pnictogen.\cite{growth,cmr,transport} These compounds exhibit a wide range of behavior, including  ferromagnetism, anti-ferromagnetism, Jahn-Teller effect, and colossal magneto-resistance, and can be found in both metallic and semiconducting phases. Of particular interest in 14-1-11 compounds are the MnPn$_{4}$ tetrahedra, which are distorted due to the Jahn-Teller effect.\cite{growth} The magnetic behavior of these materials generally reveals the Mn in a d$^{4}~3+$ state, consistent with what is expected from charge-balance. However, recent  LDA calculations and X-ray magnetic circular dichroism (XMCD) work suggests the Mn are in a d$^{5}+hole$ configuration.\cite{xmcd,LDA} The XMCD work on Yb$_{14}$MnSb$_{11}$ also showed that in this ferromagnet ($T_{C}=53~$K), the Yb has a f$^{14}$ valence and thus the moment results from the MnSb$_{4}$ tetrahedra. 

\begin{figure}
\includegraphics{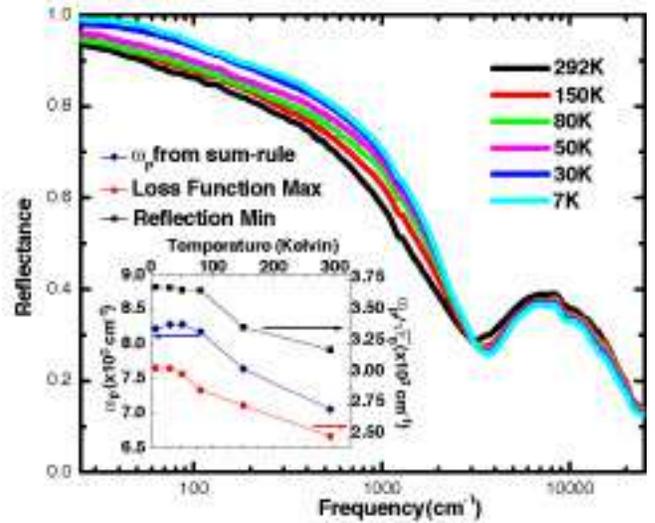}
\caption{\label{fig:ref} (color online) R$(\omega)$ of Yb$_{14}$MnSb$_{11}$ at six temperatures. Inset:  $\omega_p(T)(\widetilde{\omega_p}(T))$ as described in the text. }
\end{figure}

The Yb$_{14}$MnSb$_{11}$ single crystals were grown at ORNL as described elsewhere.\cite{transport}  This compound has been well characterized by X-ray diffraction,  magnetic susceptibility ($\chi$),\cite{growth} electrical resistivity ($\rho(T)$), specific-heat \cite{transport} and XMCD\cite{xmcd}. The sample was prepared for optical measurements by mechanical polishing, resulting in a mirror like surface. Resistivity, specific heat, magnetic susceptibility, ellipsometry and near-normal incidence reflectivity (R($\omega$)) experiments on the same single crystal were made at UCSD. The reflectivity was measured over broad frequency (25 to 25,000 cm$^{-1}$) and temperature (7 K to 292 K) ranges. The complex optical conductivity ($\sigma(\omega)=\sigma_{1}(\omega)+ i\sigma_{2}(\omega)$) was determined through a Kramers-Kronig analysis from the reflectivity measurements and confirmed at room temperature by spectroscopic ellipsometry (5,000 to 25,000 cm$^{-1}$). For the low $\omega$ region, we employed the Hagen-Rubens formula. 

In Fig. \ref{fig:ref} we plot R($\omega$) from 292 K to 7 K.  We first focus on the gross features seen in the 292 K spectrum, begining with the gradual decrease towards a local minimum at $\omega\approx 3100~cm^{-1}$. This response is characteristic of intra-band transitions, with the minimum signifying the screened plasma frequency ($\widetilde{\omega_{p}} = \frac{\omega_{p}}{\sqrt{\epsilon_{\infty}}} $), where $\omega_p$ is the plasma frequency and $\epsilon_{\infty}$ the high frequency dielectric constant. The plasma frequency ($\omega_{p}^{2}=\frac{ne^2}{m_{b}}$), is a measure of the kinetic energy of the carriers as n is the carrier density, $m_b$ the band mass, and e the electron charge.  The  resonance  at $\omega \approx 9,000~cm^{-1}$ results from inter-band transitions. As the temperature is reduced the kinetic energy of the free carriers is enhanced, signified by the increase in R($\omega<\omega_p$) and the blue shift of $\widetilde{\omega_{p}} $. Finally, the slope of the reflectance just below $\widetilde{\omega_{p}}$ grows dramatically, while R($\omega<1000~cm^{-1}$) flattens. 

\begin{table}

\caption{\label{TBL} Parameters determined through the combination of sum rules for the optical conductivity, magnetic susceptibility, resistivity and specific heat as described in the text. We note that the value of the Sommerfeld coefficient was determined via measurements down to 2K
yielding a value larger than previously reported in ref. \cite{transport}.}
\begin{ruledtabular}
 
 \begin{tabular}[b]{ccc|cc}
$n$&
$1.4 \times 10^{21} cm^{-3}$&
&
k$_{F}$&
0.35$\AA^{-1}$\\
$m^{*}$&
17.6 $m_{e}$&
&
$m_{B}$&
1.76 $m_{e}$\\
$E_{F}$&
0.261 eV&
&
$E_{F}^{*}$&
0.026 eV\\
$\frac{d\rho}{dT}$&
$2.1 \mu \Omega cm K^{-1}$&
&
$\frac{d^{2}\rho}{dT^{2}}$&
$0.5 \mu \Omega cm K^{-2}$\\
$\gamma$& 
145 $mJ/mol K^{2}$&
&
$T_{C}$&
49.22 K\\
$\chi_{0}$&
$0.005 cm^{3}/mol$&
&
$\mu_{eff}$&
$5.08 \mu_{B}$\\
\end{tabular}

\end{ruledtabular}

\end{table}

\begin{figure}
\includegraphics{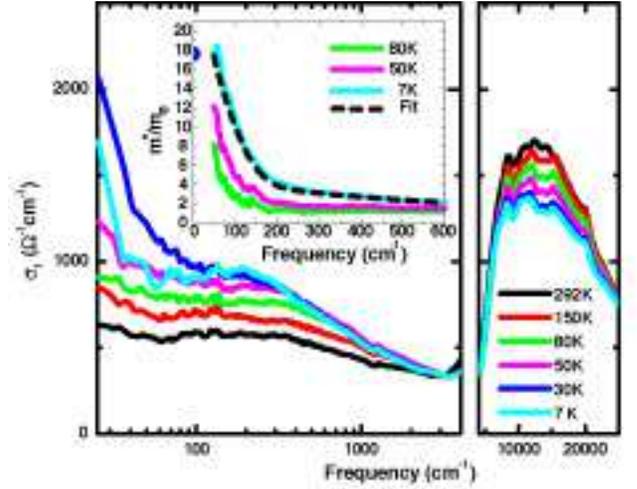}
\caption{\label{fig:s1} (color online) $\sigma_1(\omega)$ of Yb$_{14}$MnSb$_{11}$ at all measured temperatures. The clear development of a low temperature "coherent" peak can be seen. Inset: (solid/dashed lines) $\frac{m^{*}(\omega)}{m_{e}}$ from the extended Drude analysis and the fit  using Fermi's Golden rule and the density of states in Fig. \ref{fig:scale}b, (blue circle) from the specific heat analysis as described in the text.}
\end{figure}

These features are manifested in the Kramers-Kronig generated $\sigma_{1}(\omega)$ shown in Fig. \ref{fig:s1}. At 292 K, $\sigma_{1}(\omega)$ reveals two overlapping broad features extending to $4,000~cm^{-1}$. As the temperature is lowered these features develop into two distinct components. In the far-infrared region, there is a narrow Drude-like response signifying the "coherent" behavior of the free carriers,  while a mid-infrared resonance appears resulting from the "incoherent" response. The spectra, and their temperature dependence, bear a striking resemblance to what has been observed in Kondo lattice compounds\cite{sasa,hfoptics}. In particular, at low temperatures ($T<T^{*}$) a Fermi liquid with heavy quasi-particles forms resulting in a narrow Drude-like feature at low energies and a gap-like feature in the mid-infrared. As the temperature is raised above the coherence temperature ($T>T^{*}$), the carriers are no longer scattered with the same phase at each magnetic ion. Therefore, coherence is lost above $T^{*}$ and the sample reveals broad spectral features similar to an ordinary metal. Surprisingly, the coherent state in Yb$_{14}$MnSb$_{11}$ appears  to develop at the same temperature as the ferromangetism (i.e.:~$T^{*}\approx T_{C}$).

Another notable feature of $\sigma_{1}(\omega\leq4000~cm^{-1})$ in Yb$_{14}$MnSb$_{11}$ is the temperature dependence of its overall strength. To quantify the change in spectral weight we explore the temperature dependence of the plasma frequency determined through a number of means in the inset of Fig. \ref{fig:ref}. Since the reflectivity minimum and the maximum of the loss function ($-Im[\frac{1}{\epsilon}]$, where $\epsilon$ is the dielectric function) only approximate $\widetilde{\omega_{p}}$, we have exploited the well-known f-sum rule to determine the plasma frequency: $\omega^{2}_{p} = \frac{2}{\pi} \int^{\infty}_{0} \sigma_{1}(\omega)d\omega$.\cite{cut} While similar changes in spectral weight have been seen in other carrier mediated ferromagnets, they are always observed at temperatures below the onset of ferromagnetism.\cite{jasonpapers,eub} On the other hand, manganites have shown similar shifts in $\omega_{p}$ due to the competion between Jahn-Teller and double exchange interactions.\cite{manginites} However, the increase in $\omega_{p}$ seen in Fig. \ref{fig:ref} is consistent with recent Hall data where the n increases as the temperature is lowered.\cite{hall} Interestingly, the changes in $\omega_{p}$ subside for  $T\leq T_{C}$ just as the correlated state develops. As discussed later, this results from an interplay between the Kondo and Jahn-Teller effects. 

\begin{figure}
\includegraphics{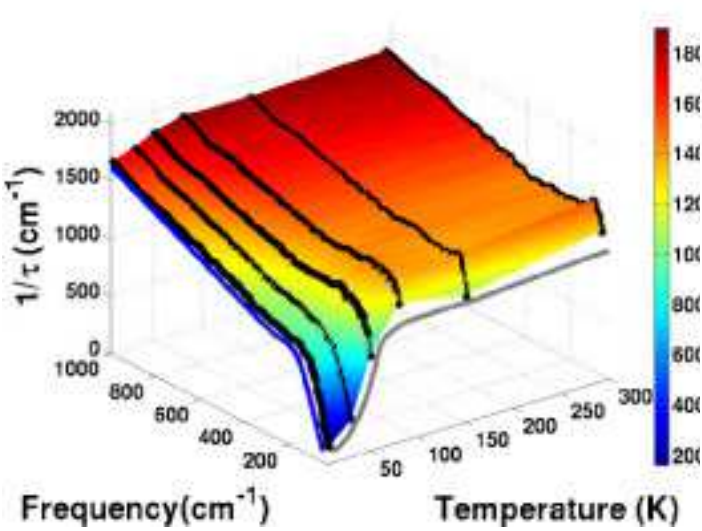}
\caption{\label{fig:tau} (color online) (black) $\frac{1}{\tau(\omega,T)}$ data with an interpolated surface, see eq. \ref{eq:tau}, (grey) the scattering rate from the d.c. resistivity ($\rho(T)=\frac{1}{\omega^{2}_{p}(T)\tau(T)}$), and (blue line) fit to $\frac{1}{\tau(\omega,7K)}$.}
\end{figure}

The coherent state in Yb$_{14}$MnSb$_{11}$ is also revealed by the quasiparticle lifetime ($\tau(\omega)$) and renormalized mass ($\frac{m^{*}(\omega)}{m_{B}}$) determined in the extended Drude formalism:
\begin{equation}
 \label{eq:mass}\frac{m^{*}(\omega)}{m_{b}}=\frac{\omega_{p}^{2}}{4\pi}\frac{\sigma_{2}(\omega)}{\sigma_{1}^{2}(\omega)+\sigma_{2}^{2}(\omega)}\frac{1}{\omega} 
\end{equation}
\begin{equation}
 \label{eq:tau}\frac{1}{\tau(\omega)}=\frac{\omega_{p}^{2}}{4\pi}\frac{\sigma_{1}(\omega)}{\sigma_{1}^{2}(\omega)+\sigma_{2}^{2}(\omega)} 
\end{equation}
In Fig. \ref{fig:tau}, we plot $\frac{1}{\tau(\omega,T)}$ demonstrating the development of long-lived quasi-particles at low energies, probed in either the temperature or frequency domains. This sudden reduction in scattering has also been seen in Kondo lattice compounds and is generally attributed to the coherent scattering of the quasi-particles by the magnetic impurities. In particular, this coherent response is believed to be a key signature of the formation of a Kondo resonance in the lattice.\cite{sasa,hfoptics,cox,coleman}  Further evidence for a Kondo resonance in Yb$_{14}$MnSb$_{11}$ is provided by $\frac{m*(\omega)}{m_{e}}$ plotted in the inset of Fig. \ref{fig:s1}. As $T$ and $\omega$ are lowered, $\frac{m^{*}(\omega)}{m_{e}}$ of the quasi-particles is enhanced, suggesting a renormalzation of the states at the Fermi energy (E$_{F}$). 

To confirm this mass enhancement, we have determined the linear term in the low temperature specific heat ($\gamma \propto n^{1/3}m^{*}$, not shown)  and renormalized plasma frequency  ($(\omega^{*}_{p})^{2} = \frac{2}{\pi}\int^{\omega_{c}}_{0} \sigma_{1}(\omega)d\omega \propto \frac{n}{m^{*}}$)\cite{wpstar} from which we can calculate the values of $n$ and the effective mass(see Table \ref{TBL}).\cite{hfoptics} The value of $m^{*}(\omega\rightarrow0)$ determined through the combination of $\gamma$ and $(\omega^{*}_{p})^{2}$ is plotted in the inset of Fig. \ref{fig:s1} and has an excellent agreement with the result of the extended Drude analysis. The $n$ determined this way is not plagued by the same difficulties as measuring the Hall coefficient and suggests that optics and specific heat measurements are excellent complementary methods to determine the carrier density in magnetic materials. This analysis also demonstrates the validity of the extended Drude analysis since we find that $(\omega_{p}(300K)/\omega_{p}^{*}(7K))^{2}=\tau(7K)/\tau(300K)=m^{*}(0)/m_{B}$.

\begin{figure}
\includegraphics{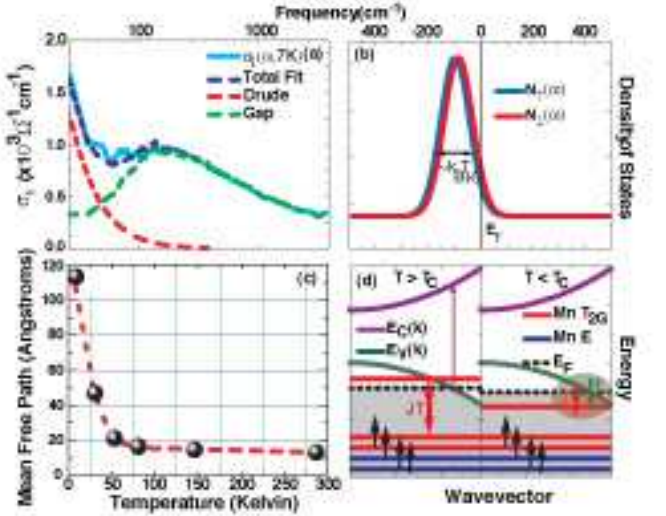}
\caption{\label{fig:scale} (color online) (a)$\sigma_{1}(\omega,7 K)$ with a Drude and hybridization gap fit.\cite{sasa} (b) $N(\omega)_{\uparrow,\downarrow}$ that fits the measured $\frac{1}{\tau(\omega,7 K)}$, as described in the text. (c) Mean Free path determined from d.c. resistivity $\rho(T)=\frac{3\pi^{2}\hbar}{e^{2}k_{F}^{2}(T)l(T)}$. (d)  Diagram of the temperature dependence of the electronic structure of Yb$_{14}$MnSb$_{11}$ consistent with our measurements.}
\end{figure}

We return to the low T/$\omega$ suppression of $\frac{1}{\tau(\omega,T)}$ and enhancement of $\frac{m^{*}(\omega,T)}{m_{e}}$ to better understand the coherent state. One can model the scattering rate in a ferromagnet using Fermi's Golden Rule for spin flip scattering: 
\begin{eqnarray}
\lefteqn{\label{eq:spinflip}
	\frac{1}{\tau(\omega)}\propto \int_{-\omega/2}^{\omega/2}[N_{\downarrow}(\omega'-\omega/2)N_{\uparrow}(\omega'+\omega/2)+{} }
\nonumber\\
& & {} \phantom{\tau( \propto \int_{-\omega/2}^{\omega/2}[}N_{\downarrow}(\omega'+\omega/2)N_{\uparrow}(\omega'-\omega/2)]d\omega'
	\end{eqnarray}
	where $N(\omega)_{\uparrow,\downarrow}$ is the density of states of the spin up, down channels.\cite{spinflip} The $N(\omega)_{\uparrow,\downarrow}$ that best fit the data are plotted in Fig. \ref{fig:scale}b, with the resulting $\frac{1}{\tau(\omega)}$ plotted in Fig. \ref{fig:tau} (blue line). The renormalized mass can then be determined through a Kramers-Kronig transformation of  $\frac{1}{\tau(\omega)}$,\cite{basov} the result of which is labeled fit in the inset of Fig. \ref{fig:s1}. Each channel of $N(\omega)$ contains a broad band ($bandwidth\approx 32,000~cm^{-1}$) and a sharp resonance ($width\approx200~cm^{-1}$). The broad band is consistent with the valence bandwidth predicted by LDA calculations of Ca$_{14}$MnBi$_{11}$,\cite{LDA} and therefore this band is attributed to the valence band of Yb$_{14}$MnSb$_{11}$. The key features of the resonance are its narrow width, $T$ dependence of its amplitude (note the lack of a sudden onset of $\frac{1}{\tau(\omega,T)}$ at high $T$), and its connection to the renormalization of the quasi-particles mass and scattering rate. 
	
	The features of the Kondo resonance are also reflected in the shape of $\sigma_{1}(\omega)$. As discussed earlier, at ($T<T^{*}$) the coherent peak in $\sigma_{1}(\omega)$ results from the renormalized heavy quasi-particles. We attribute the mid-infrared feature  to the formation of a hybridization gap in the density of states.\cite{sasa, hfoptics, coleman, cox} Specifically, it is believed that in Kondo lattices hybridization between the local magnetic moment and the valence(conduction) band results in a splitting of that band. If the band is less than half filled, then heavy quasi-particles are formed producing the coherent response, whereas the mid-infrared feature results from transitions across the hybridization gap. To confirm this assertion we have fit $\sigma(\omega,7$K) with a classical free carrier form (Drude) plus a calculation of the gap response that closely follows the BCS calculation (for more details see ref. \cite{sasa}). The resulting fit is shown in Fig. \ref{fig:scale}a. Interestingly, even at room temperature there is still some evidence of hybridization. We believe this is due to the Kondo temperature ($T_{K}=300~$K$>T^{*}$) as determined from the width of the resonance and $E_{F}^{*}$. Since the $\omega$ and T dependence of $\sigma_{1}(\omega,T),\frac{1}{\tau(\omega,T)}, \&~m^{*}(\omega)$ are explained by the existence of a Kondo resonance, we conclude that Yb$_{14}$MnSb$_{11}$ is a Kondo lattice compound. 

	Interestingly, in Yb$_{14}$MnSb$_{11}$ the Kondo physics appears to be connected to the ferromagnetism as $T^{*}\approx T_{C}$. Further evidence for the destruction of coherence above $T_{C}$ is provided by the mean free path (\textit{l}) shown in Fig. \ref{fig:scale}c and the d.c. scattering rate  shown in Fig. \ref{fig:tau}. Both establish that the sample has reached the Ioffe-Regel limit (\textit{l} = Mn-Mn distance) at $T_{C}$. We note that this reflects the importance of magnetic scattering in this sample. Nonetheless the d.c. resistivity ($\rho(T)$) displays a crossover to Fermi liquid behavior ($\rho(T)=\rho_{o}+AT^{2}$) below $T_{C}$. Furthermore the ratio of A and $\gamma$  satisfy the Kadowaki-Woods relation [16], $\frac{A}{\gamma^{2}}\approx 10^{-5}~\mu\Omega cmK^{2}/(mJ/mol)^{2}$, which is known to hold for a variety of strongly correlated Fermi liquids.\cite{kw}  
				
	While the data presented above can be attributed to the formation of a Kondo resonance, this seems at odds with the magnetic susceptibility data where the Mn appears in a d$^{4}$ configuration.\cite{transport,growth} Furthermore, X-ray data at room temperature suggest that the MnSb$_{4}$ tetrahedra have a Jahn-Teller distortion, which seems inconsistent with a d$^5$ + hole configuration.\cite{growth} These apparent inconsintencies may, however, be explained by a scenario involving the formation of a Kondo resonance between the carriers (holes) and one electron of the local Mn moment. In the ground-state the system gains charge-transfer energy by taking one of the electrons in the Sb valence band and placing it into the Mn d-level.  Also, the system can gain energy by distorting the tetrahedra such that the empty t$_{2G}$ level moves above E$_{F}$ (see Fig. \ref{fig:scale}d). At low $T$, these competing interactions may be resolved by forming a Kondo resonance between the hole in the Sb p-band and the fifth electron in the Mn d-level. This "underscreened" scenerio is possible as the Jahn-Teller effect breaks the degeneracy of the Mn d electrons. Furthermore we believe the Jahn-Teller effect can exist in this d$^{5}$ configuration since the Kondo resonance will likely mix states from the Sb p-band with the Mn levels. 
		
	The scenario described above, which extends the ideas originally proposed in references 8 and 11, has a number of interesting consequences. First the number of holes in the Sb valence band will depend on the extent to which Jahn-Teller effect overtakes the Kondo resonance, because the Kondo resonance essentially reduces the Jahn-Teller effect in order to transfer the extra electron in the Mn level. This implies that as the temperature is raised above the coherence temperature some free carrier spectral weight is transfered to the interband transition (See Fig. \ref{fig:s1} and inset of Fig. \ref{fig:ref}). Additionally the strong Hund's coupling between the electron in the t$_{2G}$ level and the local moment may explain why the coherence temperature appears to be the same as the ferromagnetic transition temperature. 
				
	In conclusion we have demonstrated Yb$_{14}$MnSb$_{11}$ can be classified as a  Kondo lattice in the underscreened limit. Furthermore we have suggested a scenario wherein intricate balance is struck between carrier mediated ferromagnetism, Kondo and Jahn-Teller effects, which is consistent with the data presented here as well as the apparent conflict between the Jahn-Teller distortion and d$^{5}$ ground state inferred from the XMCD results.\cite{xmcd} This scenario is also quite novel as it provides three "firsts", a d-electron system with a Kondo-resonance and ferromagnetism, Jahn-Teller and Kondo effects, and Kondo lattice in the underscreened limit. However theoritical work is clearly needed to clarify this picture as well as experimental studies of the $T$ dependence of the  Jahn-Teller distortion and Mn L$_{3}$ edge to quantitatively determine the evolution of the local Mn configuration. 
	
 This work was supported by the DOE and NSF. Oak Ridge National Laboratory is managed by UT-Battelle, LLC, for the U.S. Dept. of Energy under contract
DE-AC05-00OR22725. We are grateful for discussions with D.L. Cox, P. Coleman, G. Kotliar, A.J. Millis, and S.V. Dordevic.


\begin{thebibliography}
\bibitem{cut} We used $\widetilde{\omega_{p}}$ as the cutoff for the sum-rule.
\bibitem{wpstar} Here $\omega_{c}=100~cm^{-1}$ is chosen such that primarly the coherent response of the free carriers is included.
\end{thebibliography}
\end{document}